# Rheology, and how to stop aging. [a]

Jorge Kurchan

*Laboratoire de Physique Théorique de l'École Normale Supérieure de Lyon, 46 Allée d'Italie, F-69364, Lyon Cedex-07, France*

Recent analytical developments in glass theory are equally relevant to the understanding of anomalous rheology, with characteristic features such as the Reynolds dilatancy and the driving-power dependence of the viscosity arising naturally. A notion of effective temperature based on the fluctuation-dissipation relation can be introduced in the limit of small driving power. Within mean-field, the analogue of the Edwards compactivity can be computed, and it coincides with this effective temperature. The approach does not invoke any particular geometry for the constituents of the fluid, provided it has glassy behaviour.

## 1  Aging and Rheology.

Systems which have properties that depend on the time since preparation $t_w$ are said to 'age'. The simplest example is phase-separation of two inmiscible fluids. The fluids form domains whose size keep increasing with time, and if the system is infinite this process never stops. Another simple case is that of a dry foam, with the average bubble size growing like a power of time.

In many aging systems such as inorganic glasses, plastics, gels and spin-glasses, we do not have (or we do not know) a simple way to visualize 'what is growing', but we can still measure quantities that depend on the waiting time $t_w$. The measurements fall basically into two categories: two-time correlations (mean-squared displacement of particles or polymers, spin autocorrelations, etc); and responses (contraction of a sample at time $t_w + \tau$ following a pressure application since time $t_w$, magnetization evolution after a field has been turned on at $t_w$, etc.).

Figures 1 and 2 show the characteristic aging curves obtained: the typical relaxation time $\tau^{rel}$ does not immediately become infinite, but grows together with the waiting time. If a system has a very long but finite equilibration time $t^{equil}$ (like, for example, a supercooled liquid just above the glass transition), $\tau^{rel}$ grows with the age until it levels off at $t_w \sim t^{equil}$.

If we inject power into any of these aging systems, depending on the form of the drive there is the possibility of stabilizing the age of the sample in a power-dependent level: the younger the larger the power input. The phase-separation example is very clear: this is what we do when we shake a mixture of vinegar and oil to make the oil droplets smaller.

---

[a] Proceedings of 'Jamming and Rheology: Constrained Dynamics on Microscopic and Macroscopic Scales' ITP, Santa Barbara, 1997. The original talk can be found in http://www.itp.ucsb.edu/online/jamming2

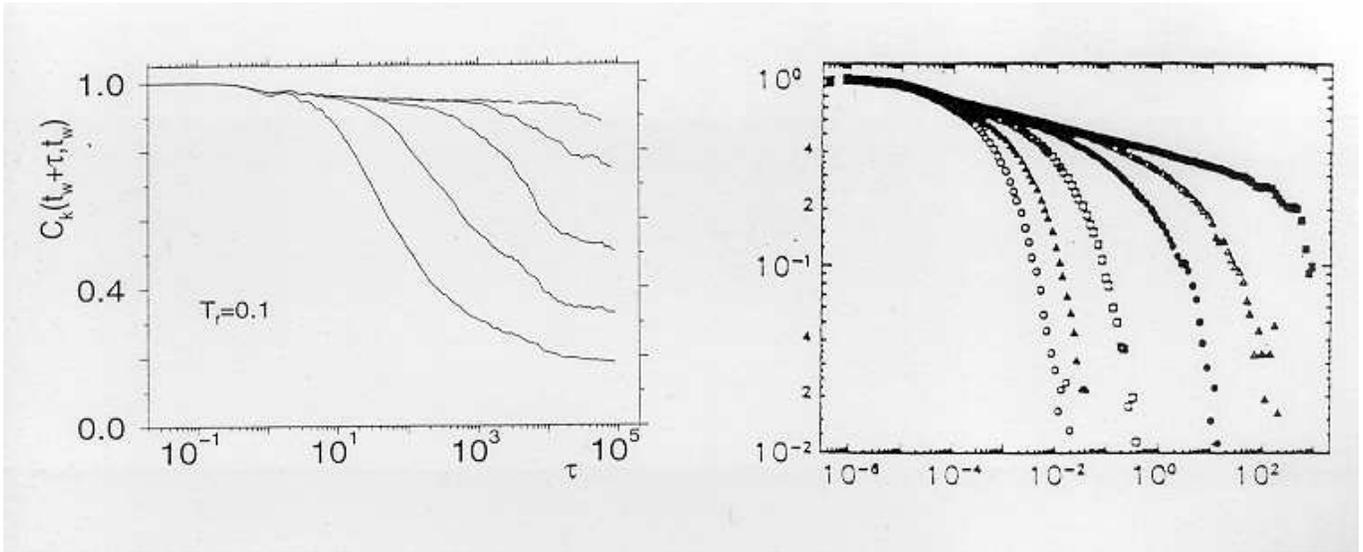

Figure 1: Autocorrelation decay for different waiting times. Left: Lennard-Jones binary mixture, molecular dynamics simulation [28] (waiting times from 10 to 39810). Right: Light scattering data for laponite gels [40] (waiting times of 11 to 100 hours). (See also [41] for similar curves for polymer melt models, [44] for spin-glass simulations, and [42,43] for polymers in random media).

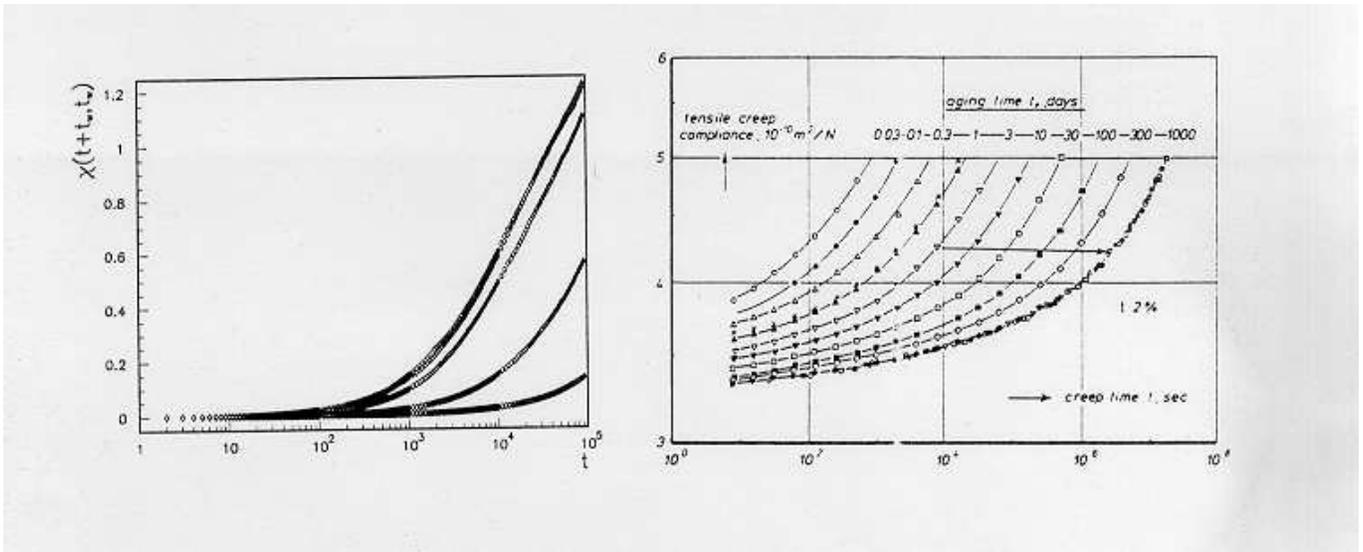

Figure 2: Response to perturbations applied after different waiting times. Left: tagged particle response in a kinetic glass model ($t_w = 10$ to $10^5$) [39]. Right: aging experiments in plastic (PVC) [36]. (See also also [37] for dielectric susceptibility measurements in glycerol, [44] for spin-glass simulations, and [43] for polymers in random media.)

The viscosity of certain gelling systems is known to increase with time, and can hence be taken as a measure of their age. When shear forces are applied to these systems their viscosity stabilises to a shear-rate dependent value [1].

Another intriguing case is that of gently, regularly tapped sand [2]. Aging under these circumstances means compactification, which makes the mobility of grains decrease with time [3]. The Reynolds effect, the fact that sand swells when sheared, is yet another example of rejuvenation and stabilization of aging through power injection.

In fact, the relation between rheology and aging in glasses is much like the relation between driven and decaying turbulence. Just like with turbulence, the driven, stationary situation is in many senses simpler.

In the last years there has been a development of analytical ideas originated in spin-glass theory. In section 2, I briefly review them as applied to purely relaxational structural glasses (there is by now quite a large literature on this case, see [4], and references therein). In section 3, I describe the implications for rheology of the same theoretical ideas. This line of research has been much less explored so far, and there are a couple of questions (Reynolds dilatancy, Edwards compactivity) that have not, to my knowledge, been discussed yet within this context.

## 2 (Relaxing) Glasses.

The considerations so far have been generic to systems with slow dynamics. It turns out that these systems fall into two classes [4]. The first class includes spin-glasses and ferromagnetic domain growth, and is characterized by having a sharp transition temperature (or density) below which slow dynamical properties appear. Systems of the second class, comprising window glasses, plastics and gels, have instead a crossover range in temperature or density in terms of which some characteristic time (typically related to the viscosity) grows rapidly. In what follows I shall only deal with the latter, although some of the discussion also applies to the former class.

*2.1 The Kirkpatrick-Thirumalai-Wolynes realization of the Adam-Gibbs scenario.*

In a series of papers [5], Kirkpatrick, Thirumalai and Wolynes pointed out that the essential features of structural glasses can be seen (albeit in a rather caricatural way) in microscopic models within an approximation that, depending on the context, is called 'direct interaction', 'mode-coupling' or 'mean-field' [7] and is exact for a family of fully connected disordered models [4]. In fact, their discussion can be extended without real complications to the whole family of approximations consisting in 'closing' the problem by reexpressing everything in terms of one and two-point correlations.

For ease of presentation, let us run the argument on the mean-field $p$-spin spherical glass[8], a model that is simple and has all the required ingredients except an ordered crystalline state (which we shall ignore completely as it is not relevant for the glass problem). The model has quenched disorder, although this is not essential, see Refs.[9]. It involves a single mode (i.e. there is no space dependence), the extension to many coupled spatial modes is straightforward[16].

This model has exponentially many stable states at low temperatures. Fig. 3 shows a sketch of the free energies of these states in terms of temperature[10]. At given temperature, the density of states per unit of free energy increases exponentially with the free energy, up until the level labeled 'threshold'. At free energies just above the threshold, the landscape has no more minima, just saddles.

The thin dashed line labeled 'liquid' corresponds to the solution that dominates at all temperatures above $T_c$ (in the language of replica theory [11] it is replica symmetric). KTW noted that the dynamical equations for the correlations just above $T_c$ are exactly the simplest[12] mode-coupling equations[13] for a liquid, with 'mode coupling transition temperature' = $T_c$. As $T_c$ is approached from above, the dynamics becomes slower and slower, the typical ('alpha') relaxation time $\tau^{rel}$ diverges.

On the other hand, an equilibrium calculation gives a transition at $T_k$, not at $T_c$. $T_k$ is for this model the Kauzmann temperature: at temperatures below $T_k$ we have an 'equilibrium glass' (Fig. 3) phase with the Gibbs distribution split between many states that have the lowest free energy density and are separated by infinite barriers. (The replica solution has a one-step breaking.)

The situation in the intermediate regime between $T_c$ and $T_k$ is more subtle. An equilibrium calculation in this range gives a solution for the Gibbs distribution that is the continuation of, and in every aspect similar to the liquid phase (the thick dashed line labeled 'non-ergodic liquid'). However, if we start from a temperature $T$ equilibrium configuration (chosen with the Gibbs probability distribution) at $T_k < T < T_c$ and study its dynamics we find that ergodicity is broken: the system never goes on to explore all the configurations belonging to the equilibrium state, but stays in a neighbourhood of the initial point. In other words, as we cross $T_c$, even if nothing spectacular happens from the point of view of Gibbs measure, the (single) state becomes fractured in exponentially many ergodic components.

It was well known that in realistic systems there was no true freezing at the mode-coupling transition temperature $T_c$. KTW pointed out that in a finite-dimensional system ergodicity would be restored between $T_k$ and $T_c$ by activated processes, and in fact one would only observe a crossover where the dynamics becomes slower than the experimental time at a certain (*so defined*) $T_g$ intermediate between $T_k$ and $T_c$. This will become more clear when we discuss the glassy dynamics in the next subsection.

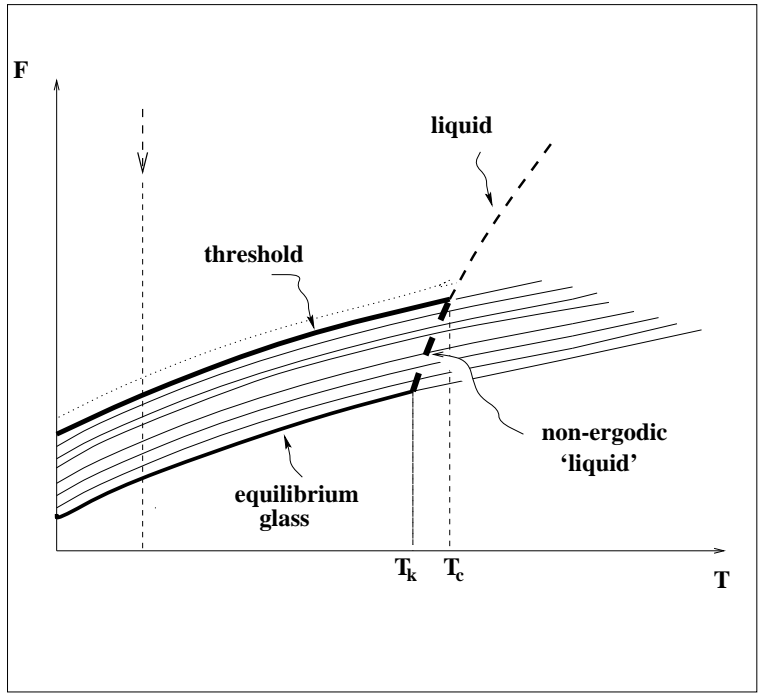

Figure 3: A sketch the temperature-dependence of metastable states

## 2.2 Out of equilibrium relaxation. Effective temperatures.

Below $T_g$, there is clearly a problem with the description above: the dynamics starting from an equilibrium configuration is invoked, although when ergodicity is broken the system could not have arrived in that configuration in the first place. In order to understand the glass phenomenon such as it happens in nature, we have to follow the dynamics of a system that undergoes a quench to low temperatures (or high pressures), and see what happens.

Within the present approximation, one has a set of two exact coupled equations for the autocorrelation $C(t,t')$ and the response function $R(t,t')$, with temperature entering as a parameter. At temperatures $T > T_c$ (the liquid phase), we find that after a short time, correlation and response functions depend only on time-differences and satisfy the fluctuation-dissipation theorem (FDT):

$$TR(t-t') = \frac{\partial C(t-t')}{\partial t'} \qquad (1)$$

indicating that the system is in equilibrium. Using (1) the two equations collapse into a single one for the correlation: these are the usual mode-coupling equations for the autocorrelations of the liquid.

If we now start decreasing slowly the temperature, we still manage to equilibrate at every step with a timescale $\tau^{rel}$ that will diverge at $T_c$. Hence, no matter how slowly we cool, there will come a temperature such that the cooling rate is too fast compared to $\tau^{rel}$. The system then falls out of equilibrium: this is signalled by the fact [14] that the correlation and response start depending on two times $C(t,t') \neq C(t-t')$ and $R(t,t') \neq R(t-t')$ and the fluctuation-dissipation relation is violated: $TR(t,t') \neq \frac{\partial C(t,t')}{\partial t'}$. Even if we stabilize the temperature at a certain $T < T_c$, neither stationarity nor FDT are achieved, no matter how long we wait. The system is *aging*: it keeps forever memory of the time since it crossed the transition. The curves of autocorrelation relaxation $C(t_w + \tau, t_w)$ and the corresponding ones for the integrated susceptibility $\chi(t_w + \tau, t_w)$ defined as:

$$\chi(t_w + \tau, t_w) = \int_{t_w}^{t_w+\tau} R(t_w + \tau, s) ds \qquad (2)$$

show the characteristic waiting-time dependence we see in Figs. 1 and 2.

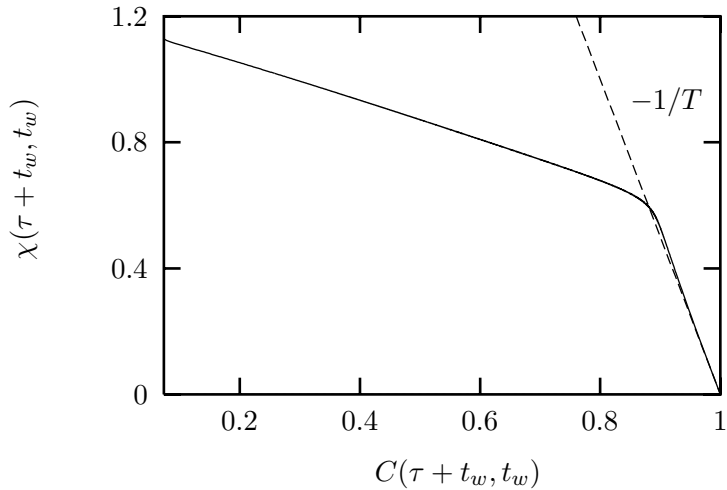

Figure 4: A fluctuation-dissipation plot (see [15,22]). The straight line to the left defines the effective temperature.

Furthermore, at each step we can compute the energy density, and find that *it is just above the threshold line* of Fig. 3. In order to understand this, bear in mind that the threshold level is such that below it there are minima in free energy, while above it only saddles: in other words, *it is the level above which the phase-space becomes connected* (hence the name 'threshold' in Ref. [14]). A system that relaxes at constant temperature, approches more and more the threshold level, thus seeing a lanscape that becomes less and less well connected, and this is why the dynamics slows down as time passes.

A parametric $\chi(t,t_w)$ vs. $C(t,t_w)$ plot[15] would give, if FDT were satisfied, a straight line with gradient $-1/T$. As mentioned above, FDT is violated in the aging regime, and we obtain a form like Fig. 4. Remarkably, for long times the plot tends to two straight lines, one with the usual gradient $-1/T$, and another with gradient (say) $-1/T_{eff}$. Thus, we have defined the effective temperature $T_{eff}$ as the factor that enters in the fluctuation-dissipation ratio.

*2.3 Beyond the simplest description.*

As mentioned above, the mode-coupling transition at $T_c$ cannot be a true one in a real life. In this context it is easy to see why: a situation with free-energy density above the equilibrium one cannot last forever, as there will always be nucleation processes in a finite-dimensional system allowing the free-energy to decrease [16].

Thus, a more realistic glass model will be able, with time, to penetrate a certain amount below the threshold level – how much depends on the thermal history. If we are at a temperature below but close to $T_c$, the system might even go down to the thick dashed line in Fig. 3, and equilibrate after a long time $t^{equil}$. Indeed, we can paraphrase Kauzmann's original argument: if we could cool slow enough, we could ideally follow all the thick dashed line, until we met the 'true' thermodynamic transition at $T_k$.

If real systems unlike the mean-field case relax below the threshold, to what extent do the aging features encountered in mean-field survive? The characteristic aging curves of Figs. 1 and 2 belong to realistic systems, and show a situation that is very similar to mean-field from the qualitative point of view. A stronger test is the existence of an effective temperature as in Fig. 4. There has been quite a lot of numerical activity to check this in finite dimensional systems, with encouraging results[27,28] (see Fig. 5). Several experiments are also now under way.

For the moment there is no real theory taking into account activated processes below $T_g$, so we have to content ourselves with learning what we can from the mean-field scenario, but always bearing in mind where are its deficiencies. This is the approach we describe next for the rheological case.

# 3 Anomalous rheology

In order to study the rheology of these models, we have to couple them to forces that can do work on them. The simplest case is to add a force field that does not derive from a global potential [17,21,18] or is time-dependent [19]. (The first studies of this kind [20] were motivated by mean-field models of neural networks, the forcing terms were added in order to destroy the undesired glass phase).

*3.1 Threshold level and Reynolds dilatancy.*

If we cool the system in the presence of a small drive down to a subcritical temperature, we can calculate the evolution of the energy. It turns out [21] that in the stationary regime obtained, the system remains 'surfing' above the threshold level, the closer the smaller the driving power: it costs an arbitrarily small power input to keep the system just above the threshold (this is, as we shall see, a mean-field peculiarity). In agreement with the discussion at the beginning, aging in this situation is interrupted, the correlation and response functions follow a typical aging pattern up to a time $\tau^{rel}$, after which they become stationary (i.e. $C(t, t') = C(t - t')$, $R(t, t') = R(t - t')$). The typical correlation $\tau^{rel}$ time depends on the strength of the non-conservative forces, and is larger the smaller these forces: hence a less driven system is more viscous, and it behaves as if it were stationarized in an older age.

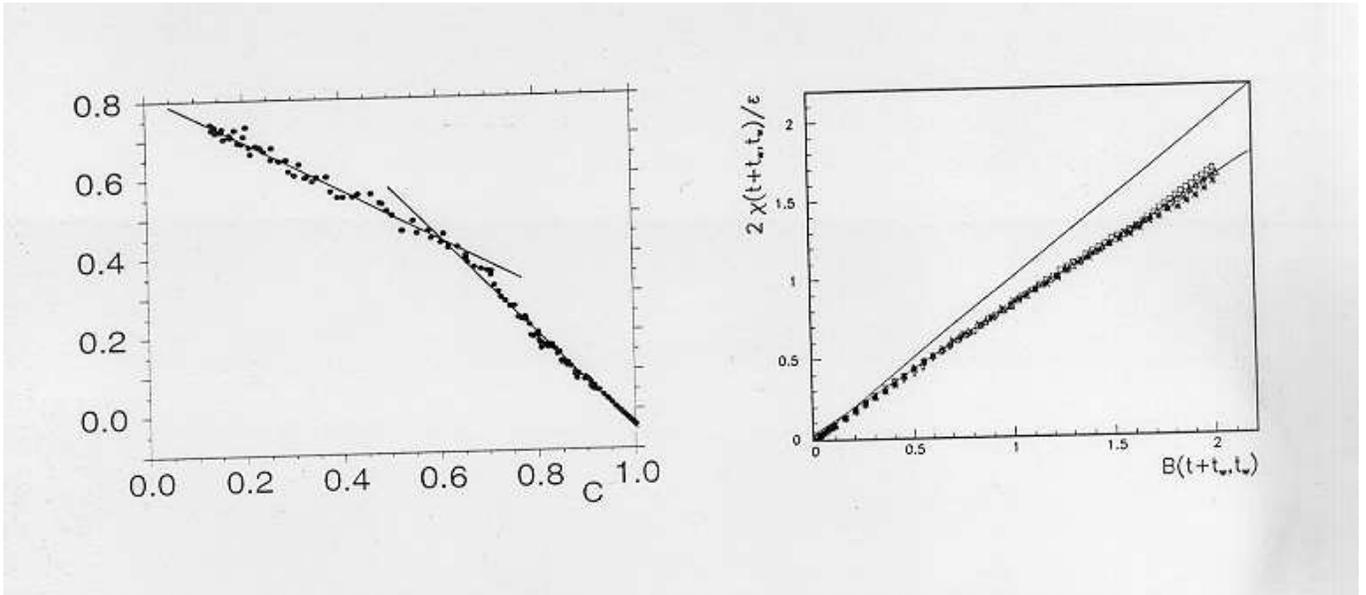

Figure 5: Fluctuation-dissipation plots. Left: Lennard-Jones glass, molecular dynamics [28]. (See [27] a similar curve for Monte Carlo simulation). Right: lattice gas [39]; the thin line is the FDT value. See [38] for the corresponding curves for ising systems, [43] for polymers in random media and [29] for models of granular media.

On the other hand, we can ask what would happen had we prepared the system in any low-lying sub-threshold state, (although this *cannot* be done within mean-field by just cooling!). For small driving forces the system remains trapped, the correlations do not decay beyond a certain small value, and the system is solid. Only by applying large driving forces can we make the system escape the deep state, and it will then resettle just above the threshold.

Hence, the system becomes free to move by raising its energy density (or volume, in the presence of gravity or pressure), the Reynolds dilatancy effect. Note that the crucial factor, the existence of a threshold level, has not been put by hand – it arises naturally from a microscopic calculation.

*3.2 Effective macroscopic temperatures. Edwards compactivity.*

When we drive a system which would age, we can make it stationary even with a very small amount of power input. We may then ask if the situation so obtained is in some sense near equilibrium. In order to test this, we make a $\chi$ vs. $C$ plot as in Fig. 4. Remarkably, in the limit of small driving the curve looks exactly like Fig. 4, not at all what we would have obtained close to equilibrium.

We have mentioned before that one can define an effective temperature $T_{eff}$ through the inverse gradient of the line to the left of Fig. 4. This corresponds to the fluctuation-dissipation ratio *associated with the slow motion*. The importance of this temperature is that one can show [22] that:

• A thermometer tuned to respond only to the lower frequencies will measure exactly $T_{eff}$.
• The effective temperatures of two different systems become the same when the systems are coupled strongly enough [b].
• The effective temperature is *macroscopic*: it stays non-zero in the limit of zero thermal bath temperature (that is, in the limit in which Boltzmann's constant is negligible).

On the other hand, Edwards and collaborators [24] introduced a definition of 'packing entropy' of a granular medium as the logarithm of the number of packings at given volume, and from it defined a 'compactivity' (playing the role of a temperature) as the inverse of the derivative of this entropy with respect to the volume. Within the present context, Edwards' 'temperature' is, at zero bath temperature, just the inverse of the logarithmic derivative of the number of stable minima of the energy with respect to the energy. (A generalization for non-zero bath temperature can be readily made substiting 'energy' by 'TAP-free energy'). Using the known [25,26] density of minima for this model, one can readily compute Edwards' temperature evaluating the energy at the threshold level, as is appropiate for weakly driven systems. *One finds that it coincides with the fluctuation-dissipation $T_{eff}$ temperature defined above*[c]. Hence, Edwards' temperature inherits the 'zero-th law' properties of $T_{eff}$ we have described.

Let us point out a suggestion that this mean-field like scenario already gives us. We have mentioned before that the plot of Fig. 4 only becomes two stright lines in the limit of small drive (or, in an aging problem, for large times). Hence, we have a concept of a single, well-defined $T_{eff}$, with the properties described above, *only in this limit*. It is then plausible the compactivity concept might itself only be relevant for weakly driven granular media, (and this quite apart from the question of its validity beyond mean-field).

An interesting question is to what extent the particular form of driving affects the stationary measure attained. This dependence should be weak, or at least controllable, if 'ergodic' arguments (as in Edwards compactivity) are to be useful. In the mean-field case

---

[b]If the coupling is too weak, something different [23] happens.
[c]This is directly checked by calculating the derivative of (2.17) in [26], and comparing with the value of $T_{eff}$ in [14]

one can easily check that observables that are not correlated with the driving force take the same values independently of the form of these forces *provided they are weak*[21]. It would be very interesting to know to what extent this is a mean-field peculiarity.

*3.3 Beyond the simplest description.*

The mean-field calculations are useful because they sometimes give results that were not expected *a priori*, and that may carry through to realistic systems: when the mean-field dynamical calculation showed the existence of an FDT temperature, this was looked for and found numerically in realistic aging models [27,28]. One can expect that the same will happen for realistic driven systems (see Ref.[29]).

In order to understand experiments and simulations it is however necessary to be aware of the limitations of the preceeding discussion, and to have an at least qualitative idea of what new elements activated processes bring in. As we have seen above, in finite dimensions nucleation allows for penetration below the threshold level [16]. This means that in order to mantain a stationary situation 'surfing' above the threshold, we need some finite (although small) energy input, unlike the mean-field situation where this could be done with arbitrarily small drive. Furthermore, the possibility of getting trapped and untrapped below the threshold can produce intermittency (see [21]) and hysteresis effects.

The greatest challenge now is the inclusion of activated, non-perturbative processes in the analytic treatment.

*3.4 Relation with other approaches*

Let us conclude by very briefly mentioning three other approaches to the rheology of soft glasses that are potentially related to the one discussed here.

A possible strategy[30,3] is to study finite dimensional systems whose infinite-dimensional version falls within the solvable category described above and that reproduce many of the qualitative features of soft glasses. Note that the scope there is not so much to find new microscopic models of glasses (polymer models, Lennard-Jones particles, etc. are already themselves perfectly good candidates), but rather models which 'interpolate' between reality and some solvable limit.

Another quite different approach is the 'soft glassy rheology' (SGR) model developed by Sollich et.al. [31]. It is built upon Bouchaud's (purely relaxational) trap model[32], supplementing it with driving forces plus – and this is essential – a *macroscopic effective temperature* representing the noise generated by the interactions. A very interesting question that arises immediately is whether the FDT-related effective temperature we discussed above is in any sense a microscopic derivation of the noise temperature in the SGR model. The question merits further consideration, but one should note that in spite of the fact that the trap models were originally inspired in mean-field statics, their dynamics is distinct and not directly related to mean-field dynamics (see [33,4] for a lengthy discussion of this point).

Another recent approach is due to Hébraud and Lequeux [34], where in particular they discuss the amplitude dependence of the response to an oscillating strain. Work is in progress[35] to study this feature within the present scenario.

## 4 Conclusion

Compared to other, more mesoscopic analytical approaches to anomalous rheology, the present one has the disadvantage that the mechanisms involved are not as explicit. It has, however, the merit of being able to give unforseen results (threshold level, FDT violations, two transition temperatures, etc.), and this because there is quite a large distance between what goes into the model (microscopic Hamiltonian and dynamics) and what comes out of it (macroscopic correlations and responses).

Acknowledgements I wish to thank Leticia Cugliandolo and Remi Monasson for discussions, and very specially Daniel Bonn for many conversations on the experimental results.